\documentclass[final,3p,times,twocolumn,authoryear]{elsarticle}

\usepackage{graphicx}

\usepackage{amssymb}
\journal{Planetary and Space Science}

\begin{document}

\begin{frontmatter}



\title{The January 7, 2015, superbolide over Romania and structural diversity of meter-sized asteroids}


\author[ond]{Ji\v{r}\'{\i} Borovi\v{c}ka}
\author[ond]{Pavel Spurn\'{y}}
\author[sarm]{Valentin I. Grigore}
\author[ta3]{J\'{a}n Svore\v{n}}

\address[ond]{Astronomical Institute of the Czech Academy of Sciences, CZ-25165 Ond\v{r}ejov, Czech Republic}
\address[sarm]{SARM - Romanian Society for Meteors and Astronomy, CP 14, OP.1, T\^{a}rgovi\c{s}te 130170, Romania}
\address[ta3]{Astronomical Institute of the Slovak Academy of Sciences, SK-05960 Tatransk\'{a} Lomnica, Slovakia}

\begin{abstract}
Superbolides, i.e.\ extremely bright meteors produced by entries of meter-sized bodies into terrestrial atmosphere, are
rare events. We present here detailed analysis of a superbolide, which occurred over Romania on January 7, 2015.
The trajectory, velocity, and orbit were determined using two all-sky photographs from a station of the European Fireball
Network (EN) in Slovakia and five casual video records from Romania, which were carefully calibrated. Bolide light curve was
measured by EN radiometers. We found that the
entry speed was $27.76 \pm 0.19$ km s$^{-1}$, significantly lower than reported by US Government sensors.
The orbit was asteroidal with low inclination and aphelion inside Jupiter's orbit. The atmospheric behavior was, however, not typical for
an asteroidal body.
The peak brightness of absolute magnitude of $-21$ was reached at a quite large height of 42.8 km and the brightness then decreased
quickly. The bolide almost disappeared at a height of 38.7 km, leaving just a stationary luminous trail visible for several seconds.
Only one small fragment continued until the height of 36 km. Brief meteorite searches were unsuccessful. 
The modeling of the light curve revealed that the body of initial mass of about 4500 kg remained almost intact 
until the dynamic pressure reached 0.9 MPa but it was then quickly disintegrated into many tiny fragments and dust under 1--3 MPa.
A comparison was made with three other superbolides for which we have radiometric light curves:
ordinary chondrite fall Ko\v{s}ice, carbonaceous chondrite fall Maribo, and cometary Taurid bolide of October 31, 2015.
The Romanian superbolide was not similar to any of these and represents probably a new type of material
with intrinsic strength of about 1 MPa.

\end{abstract}

\begin{keyword}
meteor \sep meteoroid \sep asteroid \sep atmospheric entry


\end{keyword}

\end{frontmatter}

\section{Introduction}
\label{intro}

Superbolides are extremely bright meteors, brighter in maximum 
than 
absolute (i.e.\ as seen from the distance of 100 km) 
visual magnitude of $-17$ \citep{super}. Although bolide brightness depends
on many parameters, e.g.\ meteoroid size, structure, and composition, entry speed and entry angle, we can
roughly say that superbolides are caused by meteoroids of the sizes of the order of one meter and larger. 
In fact, meter-sized bodies are now often called asteroids rather than meteoroids \citep{terminology}. 
Regardless the terminology, bodies of these sizes belong to the least known objects in the Solar System.
Current astronomical telescopes are sensitive enough to detect them when passing within few hundreds of thousands
kilometers from the Earth, nevertheless, rarely is more information obtained than the orbit. Even the size is only approximate
if albedo is not known.
Properties, and in particular internal structure, of small asteroids are, nevertheless, of great interest from several reasons.
Rotational rates of asteroids larger than about 200 meters (and smaller than about 10 km) suggest that large majority of them
are weakly bound gravitational aggregates, so called rubble piles \citep[e.g.][]{Pravec}.
Smaller asteroids rotate generally much faster and can be monolithic, although it was suggested that
many of them may be rubble piles as well, bound together by small van der Waals forces \citep{SS}.
Clarifying the question if small asteroids are monolithic or aggregates would shed light to the impact processes, in which
these bodies are born. But even monolithic materials can have various structures and strengths, depending on the
presence of cracks and other failures. Finally, even pristine Solar System materials have significant diversity as evidenced 
by meteorites, ranging from hard pure metals to relatively soft carbonaceous bodies.

Although meteorites are our best source of knowledge about microscopic and small-scale properties of asteroidal material,
they are telling only part of the story. First, meteorites represent only the strongest parts of the original meteoroid or asteroid 
--  the part, which was able to survive the atmospheric flight. Second, there are certainly meter-sized objects
that are so weak that they do not drop any meteorites. An example was the body which produced the \v{S}umava superbolide
\citep{icarus96}.

The structure of small asteroids is of interest not only from purely scientific reasons. There are ideas of future retrieval and exploitation
of small asteroids \citep[e.g.][]{capture}. The knowledge of physical properties of the target asteroid would be certainly crucial
for the success of such an attempt. Small asteroids also pose non-negligible hazard when they enter the Earth's atmosphere.
The hazard was demonstrated by the crater producing Carancas impact \citep{CarancasB, CarancasT} 
and by the damaging blast wave produced
by the Chelyabinsk entry \citep{Chelya1,Chelya2}. The actual effects on the ground caused by each particular entry depend to
large extent on the internal structure of the impactor.

Observations of superbolides can, in fact, most easily provide information about the structure of meter-sized asteroids.
Superbolides, however, are extremely rare events 
when observed from one site or from a limited region. \citet{Ceplecha94} compiled 13 bolides  
from decades long observations by three fireball networks (plus one satellite tracked bolide), which he believed
were all produced by bodies larger than 1 meter.
He classified them according to the PE criterion  developed earlier for smaller bolides \citep{PE} and based on
bolide end height. Five of the 14 bodies were classified as soft cometary (the weakest known material). 
The remaining ones were mostly classified as carbonaceous.

These results were not confirmed by the recent study of \citet{Brown}, which was based mostly on data from the US Government sensors 
obtained on global scale. The available data about 47 superbolides included only the location, height of maximum, velocity vector,
and total radiated energy. Only one event had very high maximum (59 km), suggesting its cometary nature. The others had maxima
at wide range of heights, 19 -- 45 km, without obvious grouping. While those penetrating deeply were likely compact stones or metals, 
the nature of most bodies is not possible to judge from the data. They may have been fractured objects or bodies composed from weaker
(e.g.\ carbonaceous) material.

\citet{Popova} compiled data on 13 instrumentally observed meteorite falls. They were able to compare the atmospheric behavior with
the properties of the recovered meteorites. More detailed observational data enabled the authors to find individual fragmentation
points along the bolide trajectories. For all ordinary chondrites, the inferred strength of the incoming body
was found to be one to two orders of magnitude lower than the meteorite tensile strength, presumably as a result of their collisional history.
Low strength of some carbonaceous bodies may result, on the other hand, from their porous structures, more than fractures.

\begin{figure}
\includegraphics[width=\linewidth]{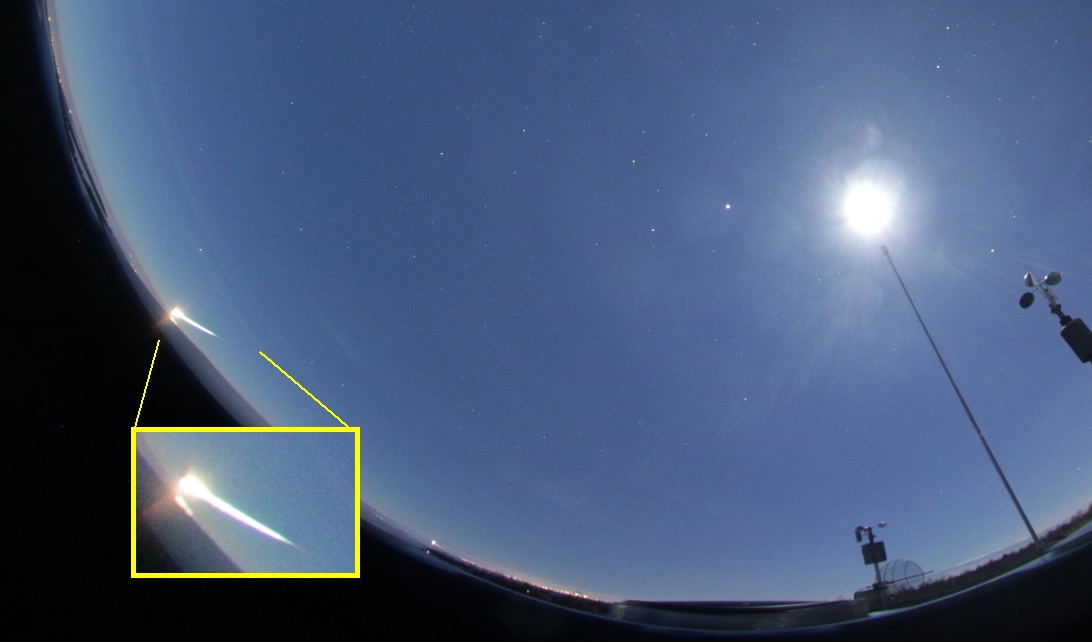} 
\caption{Part of the all-sky image taken by the Digital Autonomous Fireball Observatory in Star\'{a} Lesn\'{a} showing the January 7
bolide over the south-eastern horizon (enlarged in the inset). The brightest object on the sky is the Moon (exposed for 35 s).}
\label{Lesna-digi}
\end{figure}

\begin{figure}
\includegraphics[width=\linewidth]{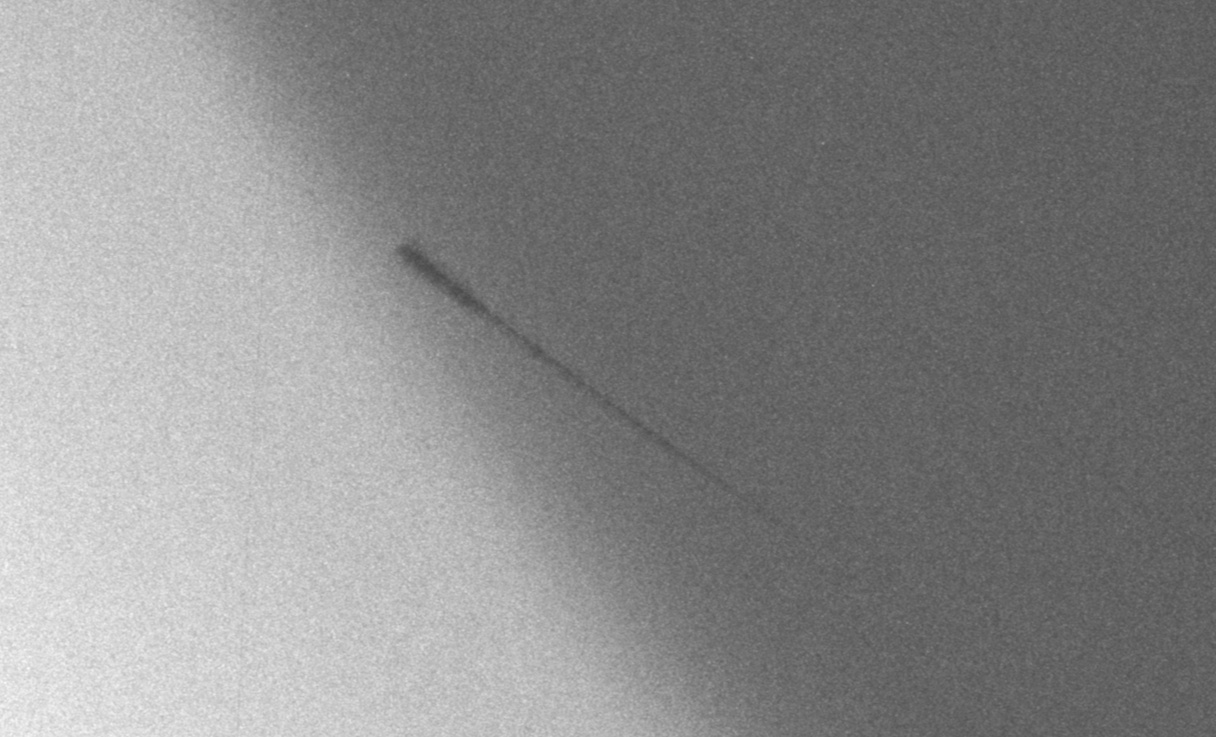} 
\caption{The January 7 bolide on the all-sky image taken by the  Autonomous Fireball Observatory in Star\'{a} Lesn\'{a} on photographic film.}
\label{Lesna-film}
\end{figure}

The study of \citet{Popova} was restricted to meteorite dropping events (of all sizes) and the study of \citet{Brown} was based on
very limited amount of data for individual events. Here we take advantage of good observational data we collected on the superbolide
which occurred over Romania on January 7, 2015. It was a very bright event caused by a meter-sized asteroid
but did not drop meteorites (at least none was found and
none larger than few grams could be expected). In addition to photographic and video records
we have also obtained detailed radiometric light curves and we could model the atmospheric
behavior of the body. The results could be compared with the modeling of three other superbolides of similar brightness, one
ordinary chondrite, one carbonaceous chondrite, and one cometary body. The January 7, 2015, superbolide was also among those 
observed by the US Government sensors and we can compare our trajectory and orbital data with those reported by  \citet{Brown}.

\begin{figure}
\includegraphics[width=\linewidth]{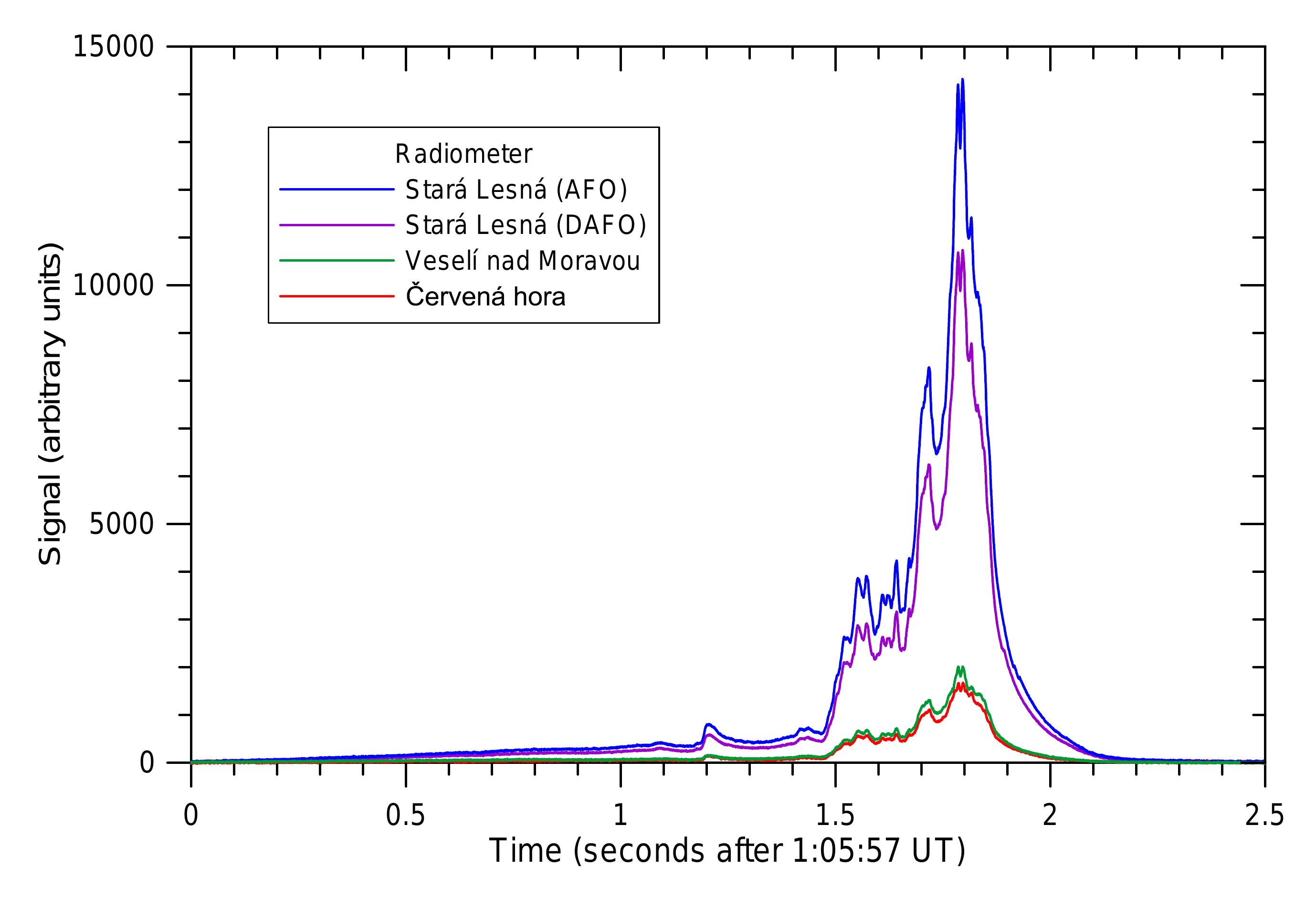} 
\caption{Light curve of the January 7 bolide in linear scale as taken by four independent radiometers. The differences in signal are caused
by differences in sensitivity and range to the bolide. The noise is about 20 units, i.e.\ smaller than the width of the lines).}
\label{radiometers}
\end{figure}

\begin{table}
\caption{Locations of various instruments used in this study}
\begin{small}
\begin{tabular}{lllll}
\hline
 Name && Longitude & Latitude & Altitude  \\
&& $^\circ$E &  $^\circ$N & m \\
\hline
 Sibiu    & V&    24.14375 & 45.78272  & 439  \\
Vo\c{s}lobeni & V&     25.63943 & 46.65971 &  825 \\
 Eforie Sud& V&   28.65293 & 44.01984 &  38 \\
 Cluj-Napoca& V&  23.61148&  46.78548 &  340 \\
 Gornovita    &V& 22.96138 & 45.06754 &  266 \\
 Star\'{a} Lesn\'{a} &AR&  20.28823 & 49.15234 &  820 \\
 Lys\'{a} hora  & A&  18.44764 & 49.54641  & 1324 \\
\v{C}erven\'{a} hora &R&      17.54196 & 49.77726  & 750 \\
 Vesel\'{\i} n.\ M. &R&      17.36962 & 48.95412 &  176 \\
 Fierbinti  &U &  26.35945 & 44.67641 & 75    \\
\hline
\end{tabular} \\
V -- video, A -- all-sky camera, R -- radiometer, U -- uncalibrated video
\end{small}
\label{coordinates}
\end{table}

\begin{figure}
\includegraphics[width=\linewidth]{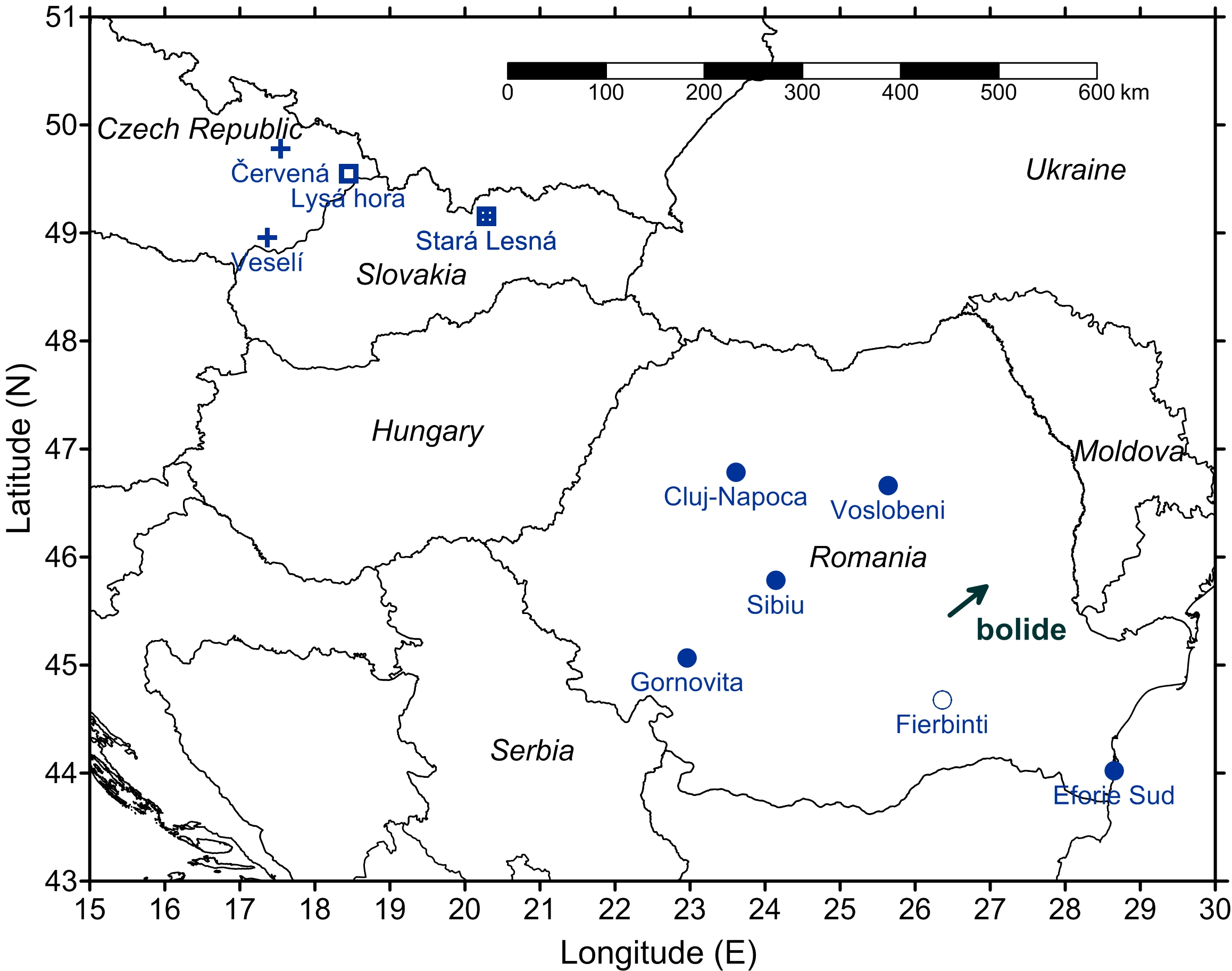} 
\caption{Map showing the location of used instruments relatively to the ground projection of the bolide trajectory. Filled circles
are calibrated videos, empty circle is uncalibrated video, empty squares are photographic all-sky fireball cameras and crosses
are radiometers (Star\'{a} Lesn\'{a} provided two all-sky images and radiometric records).}
\label{map}
\end{figure}

\section{Description of the event and available data}

The superbolide studied here occurred over central Romania on January 7, 2015, 1:06 UT (3:06 local time).
Despite the late night time, the bolide caused wide attention in the country, including news media. 
A number of videos, mostly from security cameras, showing intense illumination of the ground
lasting for about two seconds were published on the Internet. Note that bright Moon (two days after the full Moon) was present high
on the sky; the bolide was, nevertheless, much brighter.
Several video cameras, including some cameras in neighboring countries Moldova and Serbia, captured the bolide itself.
Sonic booms were reported in the region located about 100 km north of Bucharest.

The superbolide was also photographed by two all-sky cameras of the European Fireball Network (EN) located 
at Star\'{a} Lesn\'{a} in Slovakia. Although 
the bolide was more than 600 km distant from there, it was well visible close to the horizon.  Figure~\ref{Lesna-digi} shows part of the
image from the Digital Autonomous Fireball Observatory (DAFO). DAFO contains two Canon EOS 6D digital cameras equipped with
Sigma 8mm F3.5 EX DG Circular Fisheye lens. The exposure was 35 seconds long and sensitivity ISO 3200 was used. The camera contains
a LCD shutter alternating between the open and close states with the frequency of 16 Hz. Due to the high brightness and low angular
speed of the bolide, the breaks caused by the shutter are difficult to see. Nevertheless, the breaks were measurable on the original image 
(taken in the raw CR2 format) along part of the trajectory.

The second image from Star\'{a} Lesn\'{a} (Fig.~\ref{Lesna-film}) was taken by an older device, 
the Autonomous Fireball Observatory (AFO), which was
run simultaneously with DAFO. AFO uses photographic film Ilford FP125 and the 30mm fish-eye lens Zeiss Distagon F3.5 \citep{AFO}. 
The sensitivity is lower and shorter part of the bolide was recorded by this camera. On the other hand, the brightest part was not so
heavily saturated and shutter breaks could be measured, though with difficulties, along the whole recorded trajectory. Here the
shutter was mechanical with the frequency of 15 Hz.

Both DAFO and AFO are also equipped with radiometers. Both radiometers are of the same type and are based on photomultiplier
tube directed to zenith without any optics. The radiometers recorded total brightness of the sky during the bolide event with data
frequency of 5000 Hz.  The radiometers at
even more distant stations of the EN recorded the superbolide as well. Figure~\ref{radiometers} shows four independent radiometric curves. It
can be seen that all features on the light curve are well reproduced on all curves. 

Radiometers also provide exact timing
based on the GPS PPS signal. We therefore know that bolide maximum occurred at 01:05:58.793 UT (the light travel time 0.002 s to 
Star\'{a} Lesn\'{a}  was taken into account).

The bolide was also photographed by a DAFO at Lys\'{a} hora, Czech Republic, but only small part of the trajectory was captured before
the bolide crossed local horizon. 

\begin{table}
\caption{Parameters of videos and their calibrations}
\begin{small}
\begin{tabular}{llllllll}
\hline
 Name & Res & x:y & $N_{\rm ter}$  & $N_{\rm cel}$ & $\sigma$ \\
\hline
 Sibiu    & 586$\,\times\,$480 & 0.9167 & 25 & 3 & 0.12  \\
 Vo\c{s}lobeni & 800$\,\times\,$480 & 1.25 & 35 & 0 & 0.25  \\
 Eforie Sud& 704$\,\times\,$576 & 0.9167 & 37 & 0 & 0.10  \\
 Cluj-Napoca& 1280$\,\times\,$960 & 1.00 & 36 & 0 & 0.034  \\
 Gornovita    &704$\,\times\,$576 & 0.9167 & 0 & 18 & 0.06  \\
\hline
\end{tabular} 
\end{small}\\
{\footnotesize
Res -- resolution in pixels, x:y --  pixel scale ratio,  $N_{\rm ter}$ - number of terrestrial objects used for calibration,
$N_{\rm cel}$ -- number of celestial object measurements used for calibration,  $\sigma$ -- formal uncertainty
of coordinate calibration (degrees)

}
\label{videos}
\end{table}

\begin{figure}
\includegraphics[width=\linewidth]{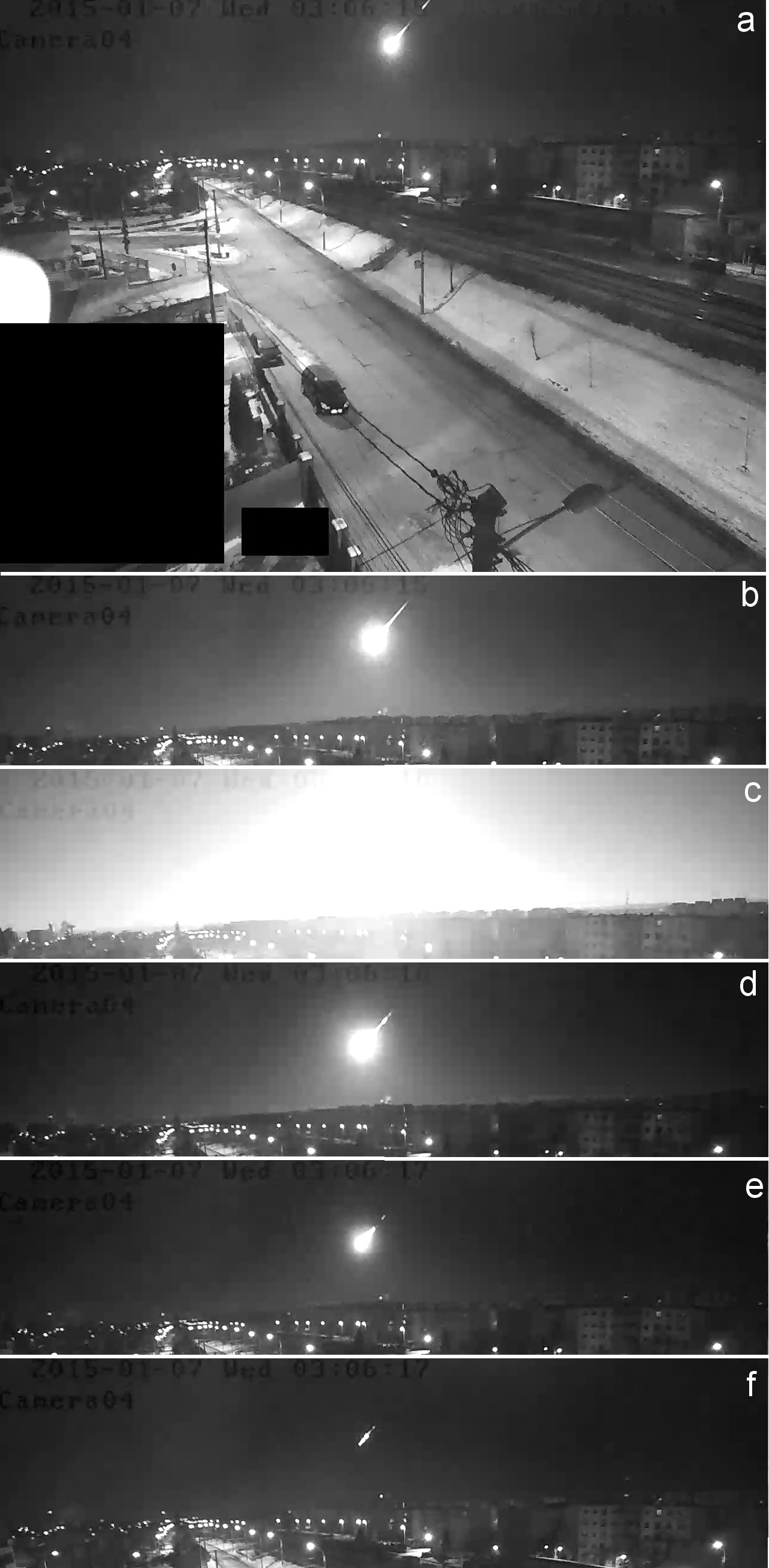} 
\caption{Selected frames from the Cluj-Napoca video: \textbf{a} -- whole frame 1, \textbf{b -- f} -- upper parts 
of frames 12, 19, 30, 42, and 58. Image c shows the bolide at maximum brightness. 
The subsequent images show just the decaying trail.}
\label{Cluj}
\end{figure}

To compute bolide trajectory and velocity we had to use also casual videos from Romania.
Casual videos need stellar calibration in order to obtain a reliable coordinate system.
Nocturnal calibration images for 4 videos were taken between January 13
and February 3, 2015. The procedure described in \citet{calibr} was used for the
calibration. All videos were taken by security cameras with known locations. In two cases (Sibiu and Eforie Sud) it was not possible to put
the calibration camera exactly at the position of the security camera. The difference in position was therefore
measured and appropriate correction was applied \citep{calibr}. 
One additional video camera (in Gornovita) was calibrated using images of the Moon and Jupiter 
recorded directly by the video camera at various times. Table ~\ref{coordinates} lists the locations of all instruments used for the
study of the bolide. It includes also another video from Fierbinti, which was taken by a dashboard camera in a moving car and could not 
be calibrated. This video, nevertheless, has the highest resolution  ($1920\times1080$ pixels, 30 frames per second)
and shows most details about the bolide.
All locations are also plotted at the map in Fig.~\ref{map}. 
The videos are provided as supplementary files to the electronic version of this paper.

\begin{figure}
\includegraphics[width=\linewidth]{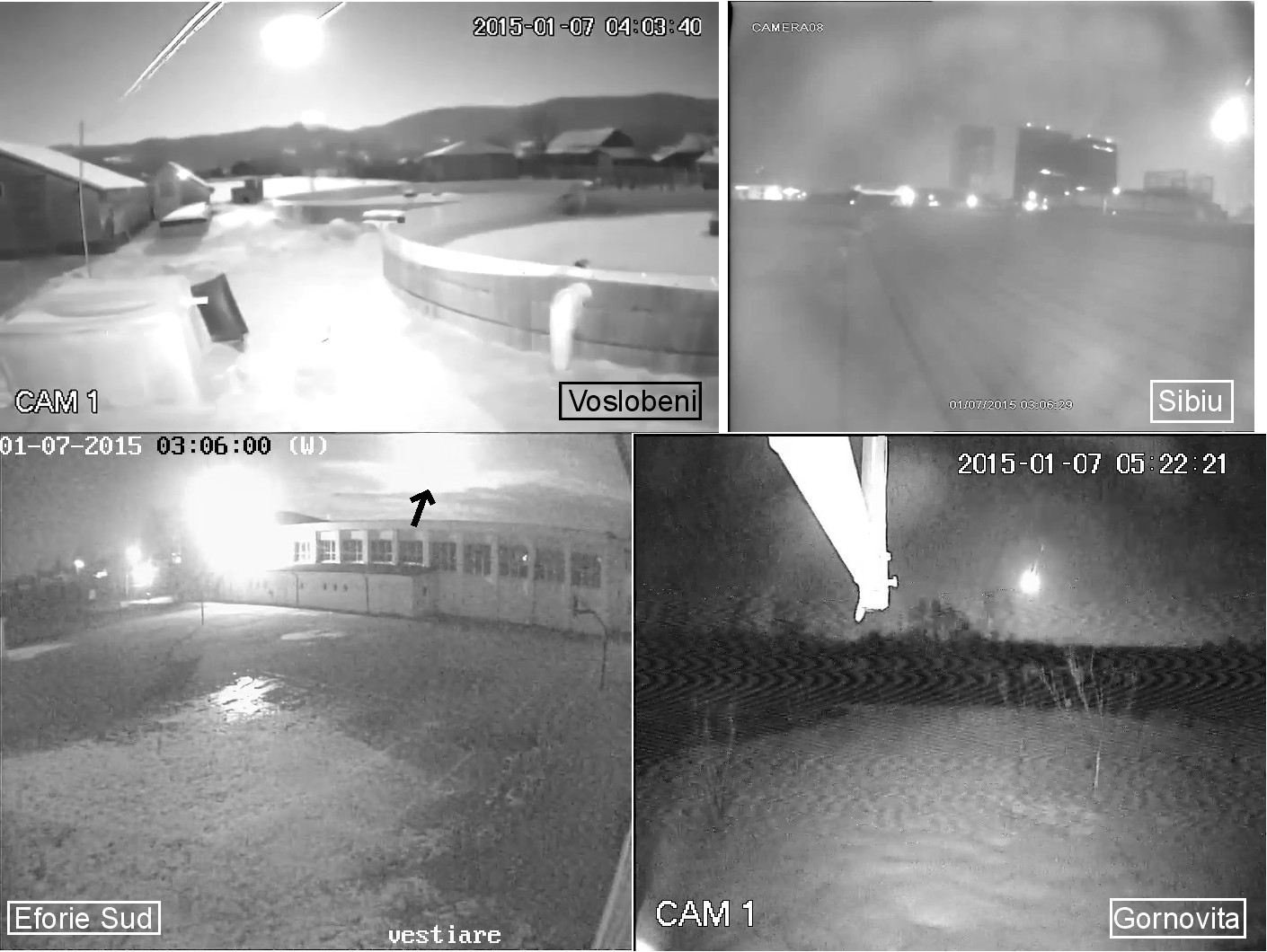} 
\caption{Whole frames from four calibrated video cameras (for Cluj-Napoca see Fig.~\protect\ref{Cluj}). All frames show the bolide in flight. }
\label{4cameras}
\end{figure}

Table~\ref{videos} summarizes the parameters of the videos and their calibration, namely the resolution in pixels, 
the pixel x:y size ratio (with one exception, the pixels were not squares), number of terrestrial objects used for calibration, 
number of celestial object measurements used for calibration, 
and the standard deviation of measurement in one coordinate. Note that the
stellar calibration images involved between 156 and 577 stars and the standard deviations were between 0.009 and 0.015 degrees.

Except Cluj-Napoca, all video had relatively low resolution. The horizontal field of view was about $110^\circ$ in all cases. All videos contained 
25 frames per second. Selected frames from the Cluj-Napoca video are shown in Fig.~\ref{Cluj}. 
The first available frame (Fig.~\ref{Cluj}a) shows the bolide in flight. Since the camera recording was activated by the bolide motion 
when the bolide became sufficiently bright, the previous part of the flight was not recorded. While the bolide path at the edge of the
field of view is well defined by the luminous trail, the velocity data are missing for this phase. In fact, 
positions of the bolide as a function of time could be measured only on frames 1--12. Later on, the bolide became too bright and
so many pixels were saturated that positional measurement was not possible.
After the maximum, only decaying luminous trail was visible for 2.5 seconds (Fig.~\ref{Cluj}d-f). 
The trail was useful for measuring bolide trajectory even during the brightest phase.

\begin{figure}
\includegraphics[width=\linewidth]{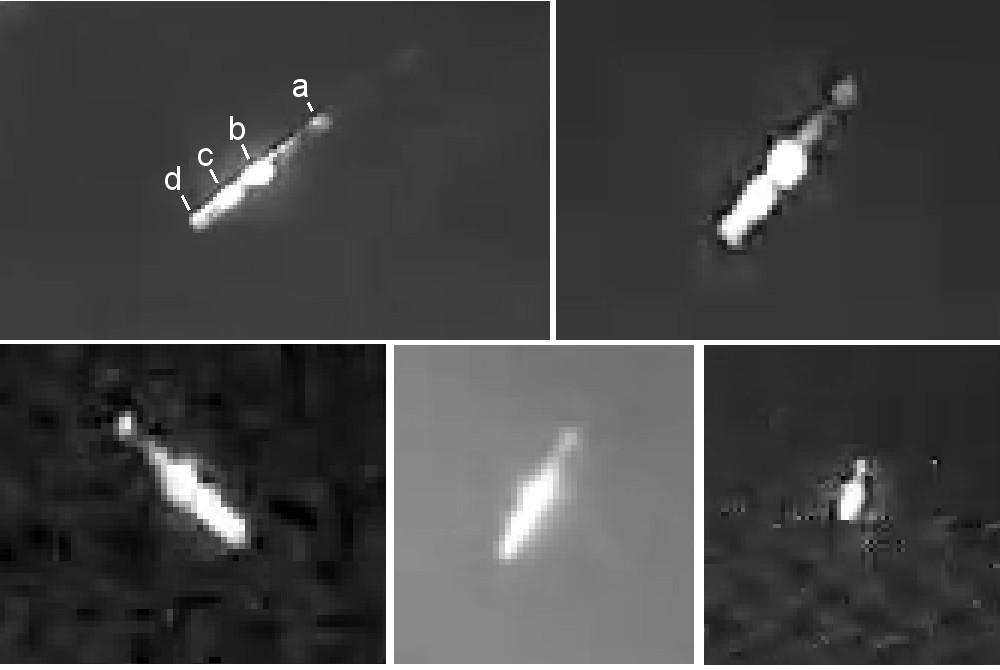} 
\caption{Detail of the trail appearance at a later stage from all five videos. Bright spots, designed a--d, could be identified in four videos.
Note that some features on some cameras were visible on other frames than shown here.}
\label{stopa}
\end{figure}

Figure~\ref{4cameras} shows video frames from other four calibrated cameras. The bolide beginning is missing in all videos. 
In Gornovita, the record started only when the bolide became bright, similarly to Cluj-Napoca. In the remaining three cases 
the bolide beginning was out of the field of view.
In Eforie Sud the bolide was already too bright to be measured when it entered
the field of view. Only the trail position could be measured and no velocity data are therefore available from Eforie Sud. 
In Sibiu the camera wobbled in the wind. Fortunately, there were only two alternative positions and the wobbling 
could be therefore easily accounted for. 

It was favorable that the brightest part of the trail exhibited, shortly before it disappeared, similar appearance in all videos.
Four bright spots were resolved on four videos (except Gornovita, where the bolide was angularly short from geometric reasons). They are
shown in Fig.~\ref{stopa}. These spots were likely present at positions where large amount of dust was deposited. 
They were useful for checking the correctness of trajectory computations. If the trajectory solution is correct,
the height of each spot in the atmosphere must be the same when computed from different cameras.

\begin{figure}
\includegraphics[width=\linewidth]{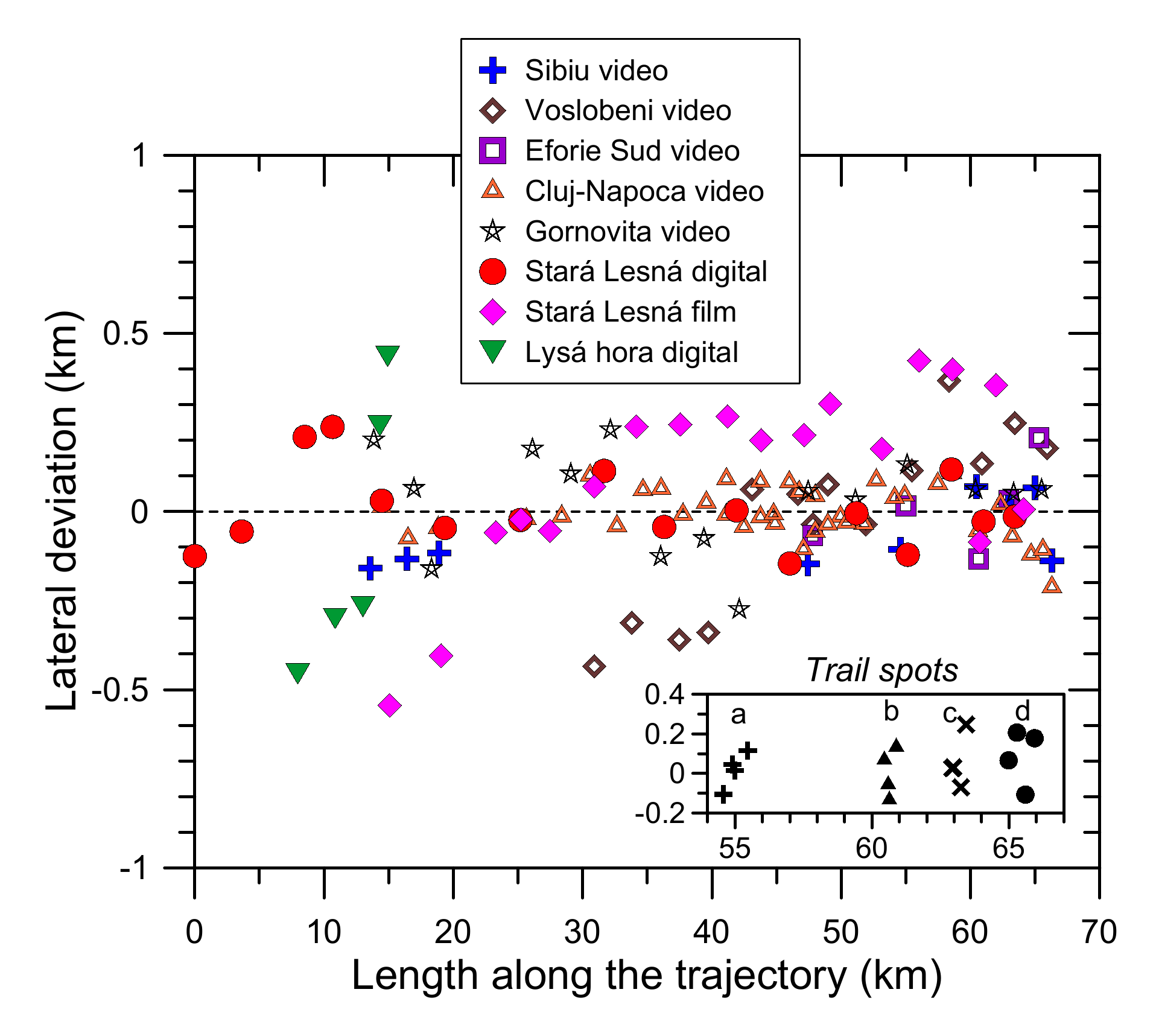} 
\caption{Deviations of all individual positional measurements from the bolide trajectory. The miss distance of the lines of sight
are given as a function of length along the trajectory (counted from the first measurement). 
Data from different cameras are given as different symbols. In the inset, the measurements of 
four bright spots shown in Fig.~\protect\ref{stopa} are given. Here different symbols represent different spots.}
\label{deviation}
\end{figure}

\section{Trajectory}

Bolide trajectory was computed by the straight least-squares method of \citet{BAC1990} using the 
azimuths and elevations of the bolide and the trail on the available videos and photographs. The method assumes that the bolide trajectory
is a straight line in the atmosphere, which was a very good approximation in this case (trajectories of long, nearly horizontally flying bolides
may be curved by gravity). The trajectory is found by minimizing the miss distances (in km) of the individual lines of sight from the
trajectory. Different weights can be given to measurements from different cameras depending on the quality of the camera and geometric
factors such as distance to the bolide and bolide angular length as seen from the camera site.
Bolide azimuths and zenith distances as measured from individual records are provided in the supplementary file \texttt{coordinates.txt}.

We succeeded to calibrate all data correctly, which is demonstrated in Fig.~\ref{deviation}. All measurements are scattered along
the trajectory with no systematic deviations and lie within 500 m of the trajectory, including data
from 630 km distant Star\'{a} Lesn\'{a} and 740 km distant Lys\'{a} hora. Only  Vo\c{s}lobeni data
show a small systematic trend. The calibration of this video was the worst (see Table~\ref{videos}) and
the video therefore got a lower weight 0.1.
The same weight was assigned to the Star\'{a} Lesn\'{a} AFO. Lys\'{a} hora, with only five measurements along
a small section of the trajectory, got a negligible weight 0.01. All other cameras had the weight 1. 

An independent check of the trajectory solution is the position of four bright spots on the trail.
Of course, since the video resolution is limited, there is some scatter. Nevertheless, as it can be seen in the inset of Fig.~\ref{deviation},
all four spots fall within 1 km in length in all four videos. We are therefore sure that the trajectory is correct. The uncertainty
in trajectory determination is about 100 meters.

The parameters of the trajectory are given in Table~\ref{trajectory}.
The beginning height of 85.5 km corresponds to the first detection
by the DAFO at Star\'{a} Lesn\'{a}. From closer distances the bolide would
be surely detected earlier. Nevertheless, as explained above, none of the Romanian videos
contain bolide beginning.
The bolide flew over the Carpathian mountains north of Bucharest, nearly from the southwest to the northeast.
The slope of the trajectory to the horizontal was $43^\circ$. The maximum energy deposition, as
documented by the brightest spot ,,b'' on the trail (see Fig.~\ref{stopa}), occurred at the height 
$42.8 \pm 0.1$ km. The heights of the other three bright spots are 46.7, 41.1, and 39.5 km, respectively, with uncertainties
of $\pm$ 0.2 km. Another distinct trail section seen on all videos was present at heights 51.5 -- 52.4 km.

\begin{table}
\caption{Trajectory and radiant of the Jan 7, 2015, superbolide.}
\begin{tabular}{lrrr}
\hline
& Longitude & Latitude & Height \\
& $^\circ$E &  $^\circ$N & km \\
\hline
Beginning & 26.450  &   45.456  &  85.5 \\
         & $\pm 0.002$ & $\pm 0.001$ & $\pm 0.1$ \\
End* & 26.948  &   45.728  &  38.69 \\
         & $\pm 0.001$ & $\pm 0.001$ & $\pm 0.07$ \\ \hline 
\end{tabular}
\begin{center}
Radiant (J2000.0)
\end{center}
\begin{tabular}{rrrr}
Right & Declination & Azimuth$^\dag$ & Zenith \\
Ascension & & (south=0$^\circ$) & Distance$^\dag$ \\ \hline
 113.8$^\circ$ &  10.13$^\circ$ & 52.2$^\circ$ & 47.0$^\circ$\\
$\pm 0.2^\circ$ & $\pm 0.08^\circ$ & $\pm 0.2^\circ$ & $\pm 0.1^\circ$ \\
\hline
\end{tabular} \\
{\small *Excluding a small fragment \\
$^\dag$At the trajectory end point}
\label{trajectory}
\end{table}

\begin{figure}
\includegraphics[width=\linewidth]{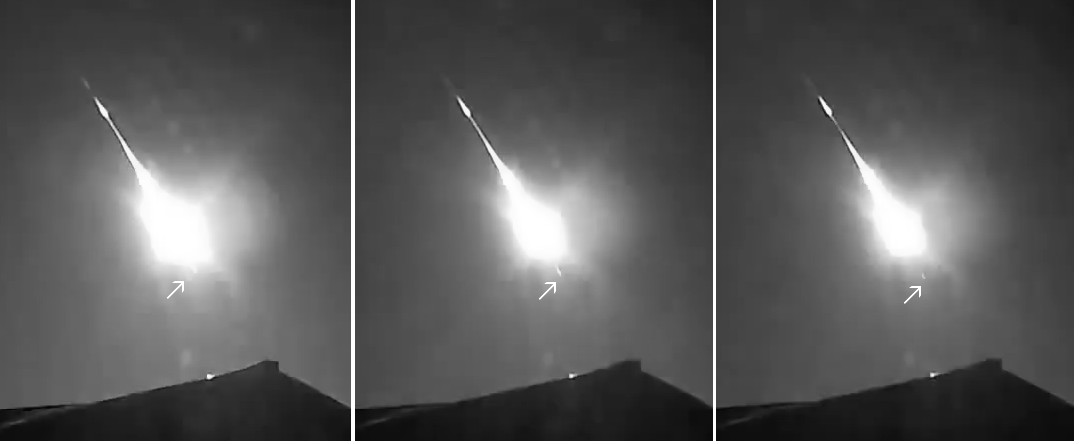} 
\caption{Part of three subsequent frames from the Fierbinti video (converted to greyscale) showing bright stationary
bolide trail and a faint moving fragment below it (marked by arrow). The fragment can be still barely seen
in the following video frame, then it disappears. In earlier frames the fragment is hidden in the trail.}
\label{fragment}
\end{figure}

The bolide end height, as seen on the calibrated videos, was 38.7 km. This is the lowest point of the freshly
visible trail. Due to the enormous brightness of the bolide in its final phase, the position of the
bolide itself could not be measured. Neither could the bolide end be measured on the Star\'{a} Lesn\'{a} digital photograph, where
it is also heavily overexposed. There is, nevertheless, a small
fragment visible in four frames of the uncalibrated Fierbinti video 
(Fig.~\ref{fragment}). At the time when the lowest
part of the trail was still very bright and saturated, the fragment 
emerged from the glow and moved further down.
Since the same bright spots in the trail were visible (at a later time) here as on the other videos,
the approximate bolide-height scale could be determined on the Fierbinti video as well.
It follows that the fragment was observed at heights 37 -- 36 km. It was also clearly visible
that fragment velocity was decreasing.

\section{Velocity and orbit}

Bolide initial velocity was determined from the measurements of bolide positions on individual video frames and
from the shutter breaks on Star\'{a} Lesn\'{a} photographs. Each position can be converted into the length
along the trajectory  and length as a function of relative time is therefore obtained. Moreover, the digital photograph
contains a time mark (one missing shutter interruption) every whole second. Combining this time mark
with radiometric light curve provides absolute timing for each shutter break. The time scales from other cameras
are then adjusted using the bolide lengths. This way all cameras are combined to obtain length as a function
of absolute time, from which the bolide velocity and, possibly, deceleration can be computed.

\begin{figure}
\includegraphics[width=\linewidth]{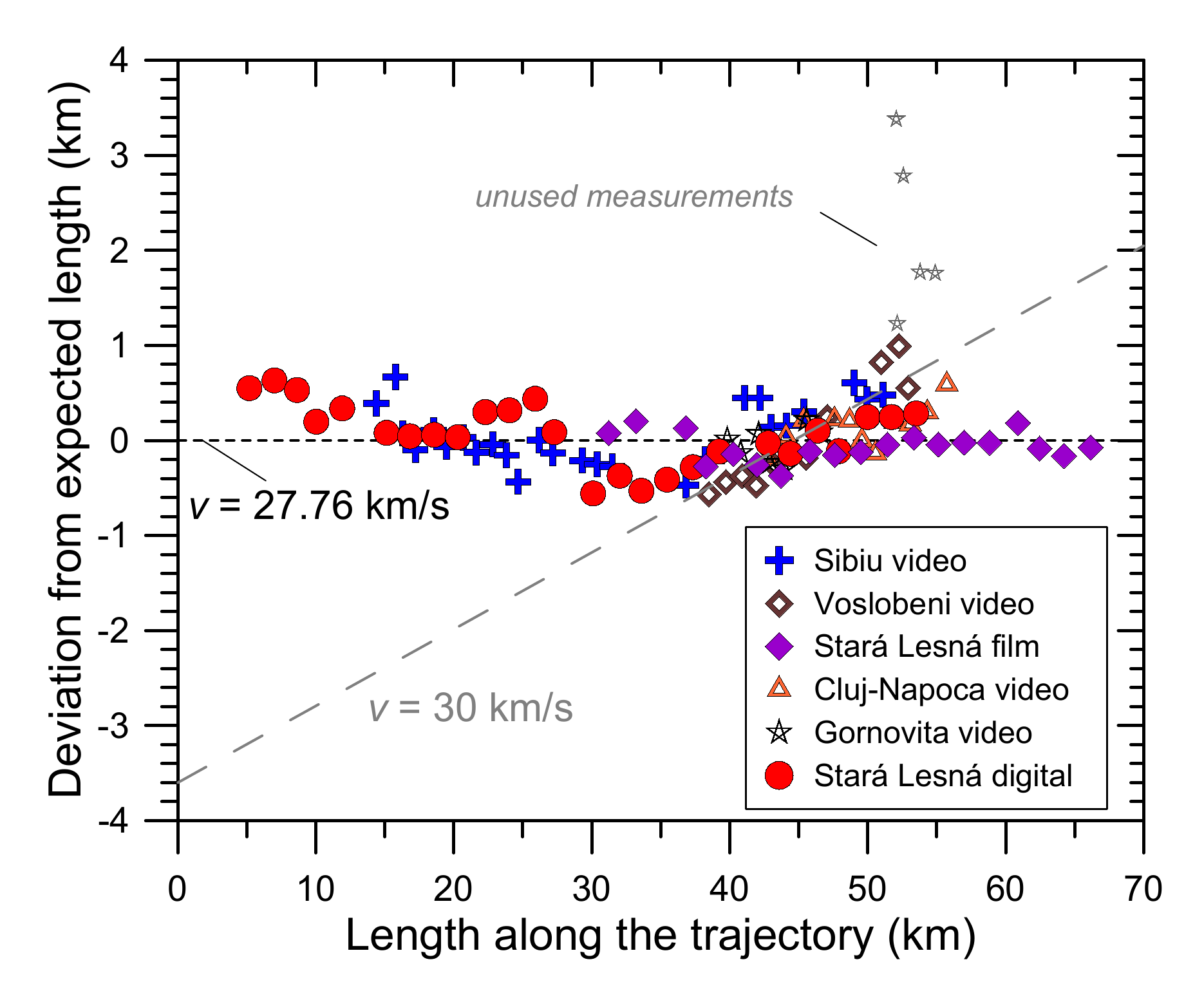} 
\caption{Deviations of the measured lengths from the length expected at the given time for a constant bolide velocity of 
27.76 km s$^{-1}$. If there were no measurement errors, all points would lie along the zero line. Outlying
measurements on Gornovita videos, which were not used for velocity determination, are shown by
small grey symbols. The long-dash line shows the expected trend for bolide velocity of 30 km s$^{-1}$. }
\label{velocity}
\end{figure}

The velocity near bolide beginning could be measured only on Star\'{a} Lesn\'{a} digital
photograph (starting from height 82 km) and Sibiu video (from 75 km). Due to its lower sensitivity,
Star\'{a} Lesn\'{a} film photograph starts at 64 km. On the other hand, the film camera is the only one which
provides velocity data during the bright phase of the bolide, down to 40 km. 
The Gornovita and Vo\c{s}lobeni videos provided really useful velocity data only
at heights 59--53 km and Cluj-Napoca at 56--48 km.  
We obtained bolide velocity $27.76 \pm 0.19$ km s$^{-1}$ 
from linear fit to the time-length data combined from all cameras.
Figure~\ref{velocity}
shows the deviations of individual length measurements from the length expected for this velocity and the given time.

Using the Vo\c{s}lobeni data only, velocity as high as 30 km s $^{-1}$ could be obtained. However, this is a result
of relatively poor calibration of this video and difficulties of measurement of the bolide in the bright phase. Cameras which recorded
larger part of the bolide, namely both cameras from  Star\'{a} Lesn\'{a} and the Sibiu video, clearly
show that the velocity was lower. Other videos also agree with this lower velocity. 
There is no sign of deceleration (within the precision of measurements) down the height of 40 km. 

Using the radiant given in Table~\ref{trajectory} and the atmospheric entry velocity of 27.76 km s$^{-1}$, the pre-impact
heliocentric orbit was computed by the method of \citet{Ceplecha87}. 
The resulting orbital elements are given in Table~\ref{orbit}. The geocentric velocity was $25.6\pm0.2$ km s$^{-1}$.
The orbit has high eccentricity and relatively low inclination. According to the Tisserand parameter, it
can be classified as an asteroidal orbit, although it is close to the boundary with Jupiter family comets at $T_{\rm Jup}=3.05$
\citep{Tancredi14}. The orbit is also close to the 7:2 resonance with Jupiter located at 2.25 AU. 
 
\begin{table}
\caption{Heliocentric orbit (J2000.0)}
\label{orbit}
\begin{tabular}{ll}
\hline
Semimajor axis &  2.27 $\pm$ 0.06 AU \\
Eccentricity  &  0.785 $\pm$ 0.006 \\
Perihelion distance & 0.489 $\pm$ 0.004 AU \\
Aphelion distance & 4.06 $\pm$ 0.12 AU \\
Inclination &  12.17$^\circ$ $\pm$ 0.11$^\circ$ \\
Argument of perihelion & 98.2$^\circ$ $\pm$ 0.5$^\circ$ \\
Longitude of the ascending node & $106.199^\circ$ \\
Tisserand parameter to Jupiter & 3.09 $\pm$ 0.06 \\
\hline
\end{tabular}
\end{table}

\section{Meteorite searches}

It was obvious from the beginning that the Romanian bolide does not represent a typical meteorite-dropping 
event. The entry speed was relatively high and the bolide end was quite sudden and occurred at
high altitude.  Nevertheless, meteorites were recovered before from superbolides of entry speeds of $\sim 28$ km s$^{-1}$
and end heights above 30 km \citep{AIV}. It was therefore not entirely hopeless to search for them also in this case.
In case of success, the meteorite-dropping-bolide end heights would be just extended to somewhat higher
altitudes. More importantly,  it would be interesting to learn what kind of material caused this unusual superbolide.

From the character of the bolide we could not expect any large meteorites. 
The asteroid was obviously completely disrupted near the bolide end.
Nevertheless, the initial mass was quite large, so there was a possibility of large number of small meteorites
(we expected sizes up to few centimeters at maximum, mostly smaller).  The possible impact area was
computed using the dark flight code of \citet{Ceplecha87} and high atmosphere winds
measured by the  Bucharest radiosonde at 0 UT. Although the bolide ended over mountains, the wind moved
possible meteorites to the flat region in the east. The most probable impact area was found to be a 2.5 km long and 0.5 km wide 
strip between 45.707$^\circ$N, 27.114$^\circ$E and  45.730$^\circ$N, 27.120$^\circ$E. Meteorite masses should increase 
in the northern direction.  Most of the area is cultivated
fields but a part of village Pietroasa, belonging to nearby C\^ ampineanca, lies also within it. 

It was a freezing weather and snow was lying on the ground during the 
event
on January 7. Shortly after that, the snow melted and
there were periods of rain and snowfall before the weather improved in the spring. 
Two expeditions of five and seven people, respectively, were organized into the suspected impact area on March 26, and on April 3, 2015. 
Searches were performed visually and with metal detector. The conditions were favorable with flat terrain and little vegetation.
Of course, only small part of the suspected area could be searched. The results were negative. The mayor of C\^ ampineanca
and several local people were interviewed. Nobody was aware of any recovered fragments.

\begin{figure}
\includegraphics[width=\linewidth]{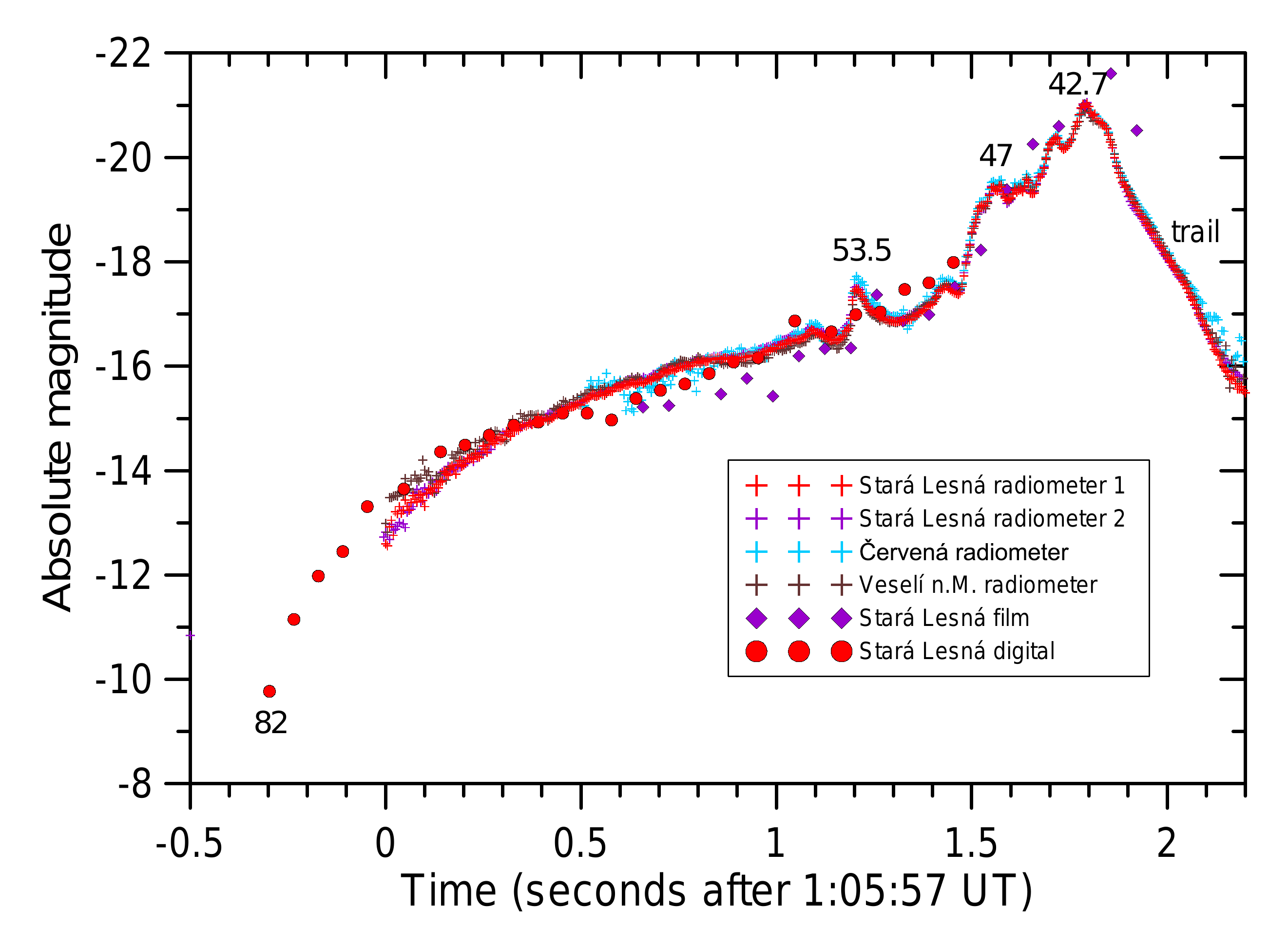} 
\caption{Light curve of the January 7 bolide determined from four radiometers, one digital photograph and one film
photograph. The numbers give bolide height in km at the corresponding time.}
\label{lightcurve}
\end{figure}

\section{Bolide light curve}

Figure~\ref{lightcurve} shows the brightness of the bolide in absolute magnitudes as a function of time. The raw
radiometric records shown in Fig.~\ref{radiometers} were converted into stellar magnitudes taking into account
the response as a function of zenith distance (measured in laboratory), the range to the bolide, and atmospheric extinction.
Since radiometers provide signal as a function of time, bolide position as a function of time must be computed to obtain
bolide range and zenith distance. For that purpose the bolide trajectory and constant velocity as determined
in previous sections were used. Note that this approach may lead to somewhat overestimated brightness of the bolide trail at the end
of the light curve since the trail did not continue the forward motion.
Radiometric curve is provided in the supplementary file \texttt{lightcurve.txt}.

The radiometers have different sensitivities and are not calibrated in absolute sense.
To obtain absolute magnitudes, radiometric curves were shifted to overlap with the light curves obtained 
from Star\'{a} Lesn\'{a} photographs. Photographic light curves were calibrated using stars. 
Their scatter is, however,
larger and temporal resolution is lower than for radiometers. Moreover, the photographs became saturated when bolide
brightness increased too much. Although saturation correction was applied, the precision in the saturated part is poor.
The digital photograph became saturated when the bolide absolute magnitude reached $-16$ and  no measurements were
possible at all above magnitude $-18$. The film camera was saturated above magnitude $-19$.

It can be seen in Fig.~\ref{lightcurve} that the agreement between independent measurements on digital and film cameras
is reasonably good. The digital camera is generally more precise. To adjust the radiometric scale, the part of the digital light
curve where the bolide brightness was about $-15$ mag was used. Here the signal was high but not saturated yet, so the uncertainty
is the smallest. 

The extinction correction was very important because the bolide was quite close to the horizon.
Note that extinction correction must be performed differently for cameras and radiometers. For cameras, we used
the formula of \citet{Rozenberg} to compute the airmass, $X$, as a function of zenith distance, $z$:
\begin{equation}
X = \big(\cos z +0.025 \exp(-11\!\cos z)\ \big)^{-1}.
\end{equation}
Radiometers, however, do not observe just the bolide. They capture all light coming to the station, including bolide light scattered
by the atmosphere. The signal therefore does not decrease so much when the object is close to horizon. This fact was discussed
by \citet{Schaefer}. Using the rising full Moon, we found the following simple formula appropriate for radiometric 
measurements at low elevations:
\begin{equation}
X_{\rm scatt} = \big(\cos z +0.1\ \big)^{-1}.
\end{equation}
The extinction correction is $\Delta m = k X$, where the
extinction coefficient  $k$ was assumed to be the same for cameras and radiometers. 
It was determined using stars on the digital camera ($k=0.25$ mag).

Although the three detectors (panchromatic film, digital camera, and photomultiplier used in the radiometers)
have slightly different spectral responses, the maximum always lies in visual range. The measured
magnitudes can be therefore regarded as visual magnitudes.

As it can be seen in Fig.~\ref{lightcurve}, the bolide light curve was smooth at the beginning. The initially quick increase 
of brightness was followed by a slower brightening. A short asymmetric flare with
steep onset occurred at the height of 53 km. A much more significant onset of brightness started at the height of 48 km. 
The rest of the bolide was a gigantic flare with several submaxima. The peak
brightness of absolute magnitude $-21.0 \pm 0.3$ occurred at the height of 42.7 km, 
in good agreement with the position of the brightest spot in the trail.
The brightness started to drop quickly at time 1:07:58.85, which corresponds to height 41.5 km, assuming no bolide deceleration. 
Most of the radiometric signal after 1:07:59 UT was due to the stationary bolide trail. The trail luminosity was
supposedly produced by a column of hot vapor, which was initially optically thick. 
Its luminosity could be also supported by continuing ablation of fine dust.

\begin{figure}
\includegraphics[width=\linewidth]{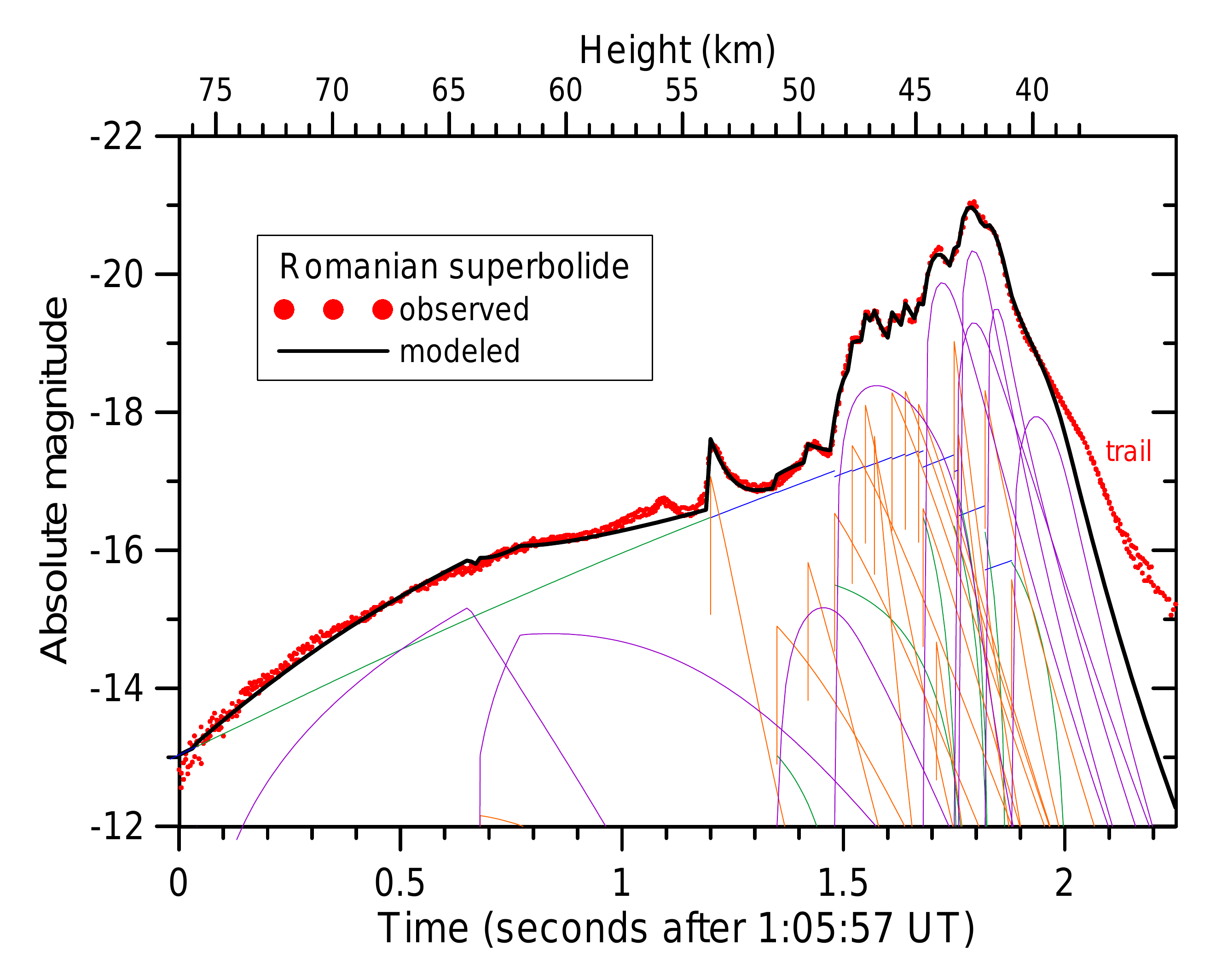} 
\caption{Observed and modeled light curve of the January 7 bolide.  Data from both Star\'{a} Lesn\'{a} radiometers
are plotted as the observed curve. The modeled curve is a sum of contributions of various fragment systems, which
are given as thin lines below the curve. Macroscopic fragments subject of partial or full erosion (release of small fragments, 
collectively called dust) are given in green. The light produced by the eroded dust is in purple. The suddenly released
dust is in orange.  Macroscopic fragments subject only to regular evaporation are in blue.}
\label{lc-model}
\end{figure}

\section{Fragmentation model}

The bolide light curve was modeled using the fragmentation model developed for the Ko\v{s}ice meteorite fall
\citep{Kosice}.  Every flare can be explained by a fragmentation event accompanied with a release and evaporation of
a number of small fragments. The complex light curve could be therefore modeled relatively easily assuming that every
increase of brightness was caused by a loss of certain amount of mass in form of small fragments. 
The fragment mass range can be estimated from the slope of the brightness onset and from the duration
of the flare. 
Within that range, the power law distribution with mass distribution index $s=2$ was assumed.
In addition to instantaneous release of fragments, gradual release was also allowed. In that case
the rate of fragment release was described by the erosion coefficient, an analogy of the ablation coefficient describing
the evaporation rate \citep[see][]{Kosice}.

The modeled light curve is shown in Fig.~\ref{lc-model}. 
Luminous efficiency was assumed to depend on fragment
velocity and mass. We used the velocity- and mass- functions of \citet{ReVelle2001}, normalized so that the luminous
efficiency at 15 km s$^{-1}$ was 5\% for large fragments ($\gg 1$ kg) and 2.2\% for small fragments ($\ll 1$ kg). The conversion factor
1500 W for zero magnitude meteor \citep{SSR} as used to convert magnitudes into 
the radiated energy (bolometric).
 The initial mass of the asteroid was estimated to be 4500 kg. Considering the uncertainty
in photometry and luminous efficiency, the uncertainty of the initial mass is about $\pm50\%$.

Atmospheric density was computed by the NRLMSISE-00 model \citep{atmos}. The material bulk density and the ablation coefficient
were assumed to be 2500 kg m$^{-3}$ and 0.01 s$^2$ km$^{-2}$, respectively. These values are different from those
used for the ordinary chondrite Ko\v{s}ice (3400 kg m$^{-3}$ and 0.005 s$^2$ km$^{-2}$), reflecting a softer nature of the Romanian body.
Ablation coefficient could not be derived from observations because no deceleration was observed.
The value $\Gamma A = 0.7$, where $\Gamma$ is the drag coefficient and $A$ is the shape coefficient, was used.

\begin{table}
\caption{Entry parameters of four superbolides compared in this study.}
\label{4bolides}
\begin{small}
\begin{tabular}{llllll}
\hline
Name & Mass$^a$ &	Speed & 	Slope &	Material & Density \\
& kg & km/s &$^\circ$ && kg/m$^3$\\ \hline
Ko\v{s}ice &	3500 &	15.0 &	60 &	H5  & 3400$^b$ \\ 
Maribo &	2300 &	28.3 &	31 &	CM &  2100$^c$ \\
Taurid	 & 1300 & 	33.1 &	17 &	Comet & 800$^d$ \\
Romanian & 4500 &	27.8 &	43 &	? & ? \\
\hline
\end{tabular}
\end{small}
\begin{footnotesize}
$^a$Photometric mass estimate;
$^b$\citet{Kohout};
$^c$\citet{Consolmagno};
$^d$\citet{Encke}

\end{footnotesize}
\end{table}

\begin{figure}
\includegraphics[width=\linewidth]{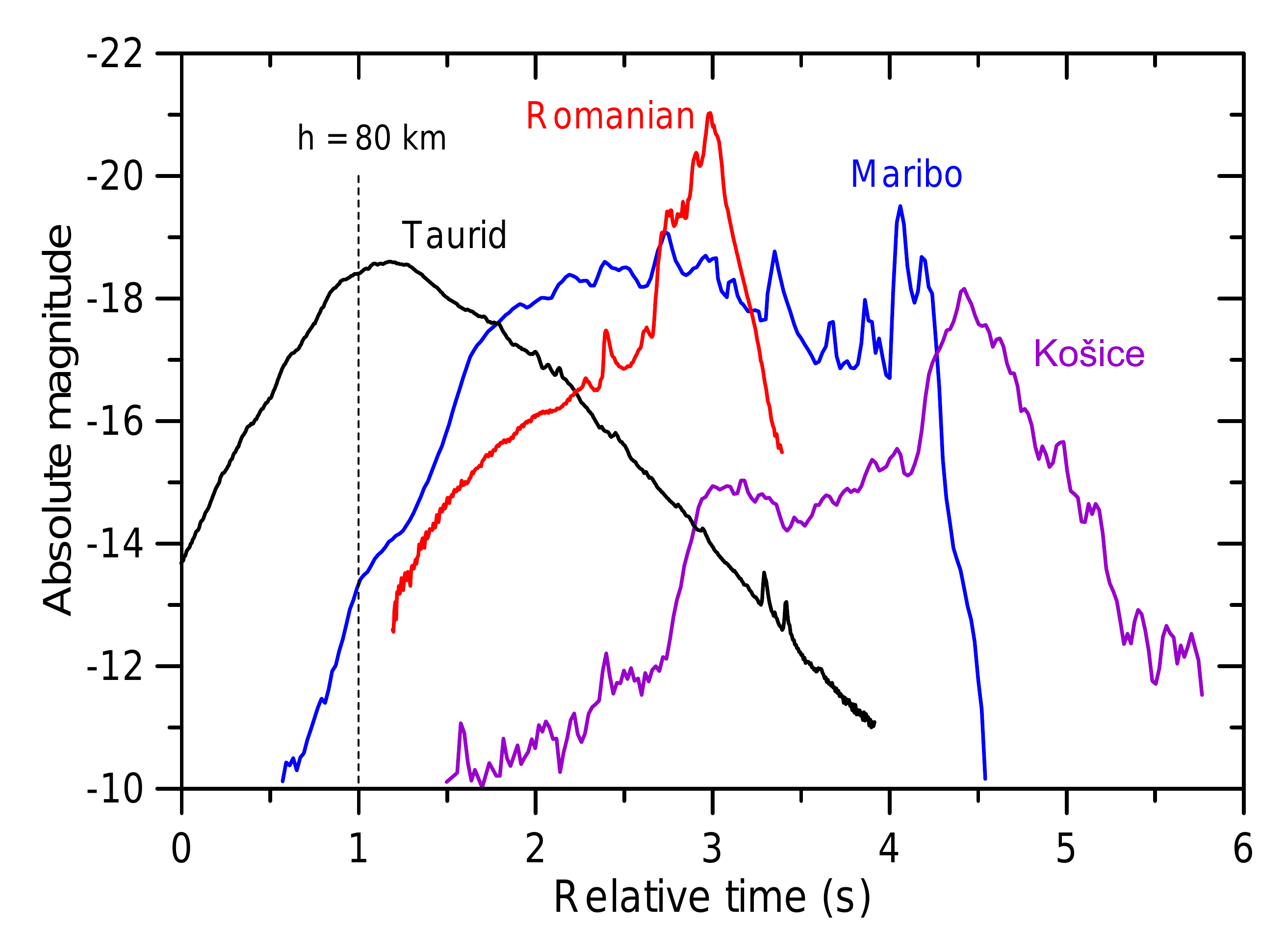} 
\caption{Calibrated radiometric light curves of four superbolides. The time scale was set so that
all bolides passed the height 80 km at time 1 second. The light curves contain trail radiation.}
\label{4curves}
\end{figure}

To explain the light curve shape, it was necessary to assume that the asteroid was gradually loosing
small fragments, presumably from surface layers, between heights 76--54 km. At the beginning, the fragments were small
($10^{-6}$ -- $10^{-5}$ kg), then they  became larger ($\sim 10^{-3}$ kg). 
All details on the light curve were not modeled in this part.
The flare at 53.5 km was 
modeled
by a sudden release of about 30 kg of mass in form of $10^{-5}$ kg fragments. 
To explain the brightest part of the bolide, more than a dozen of individual fragmentations were modeled
between heights 48 km and 42 km. In total,
850 kg of mass was lost in sudden releases and 3200 kg in form of eroding fragments.
The erosion coefficients were between 0.5 and 2 s$^2$\,km$^{-2}$.  All fragment masses were between
$10^{-4}$ and $5\times10^{-3}$ kg. We do not claim that smaller or larger fragments were not formed but they are not
necessary to explain the light curve and if they existed, they represented a minor part of the released material. 

The remaining 170 kg body started to erode at a height of 40.9 km and was almost destroyed before it reached 38 km.  
Its only remnant may be the single fragment observed to survive the main flare at the Fierbinti video. This
fragment arrived at the height of 37 km with still large velocity $\gg 20$ km s$^{-1}$. The estimated mass at that moment 
was just a few tens of grams. The subsequent quick deceleration and disappearance suggest intensive mass loss.

The model predicts only meteorites smaller than 1 gram
but we cannot exclude that some pieces larger than 1 gram reached the ground.

\section{Comparison with other superbolides}

\begin{figure}
\includegraphics[width=\linewidth]{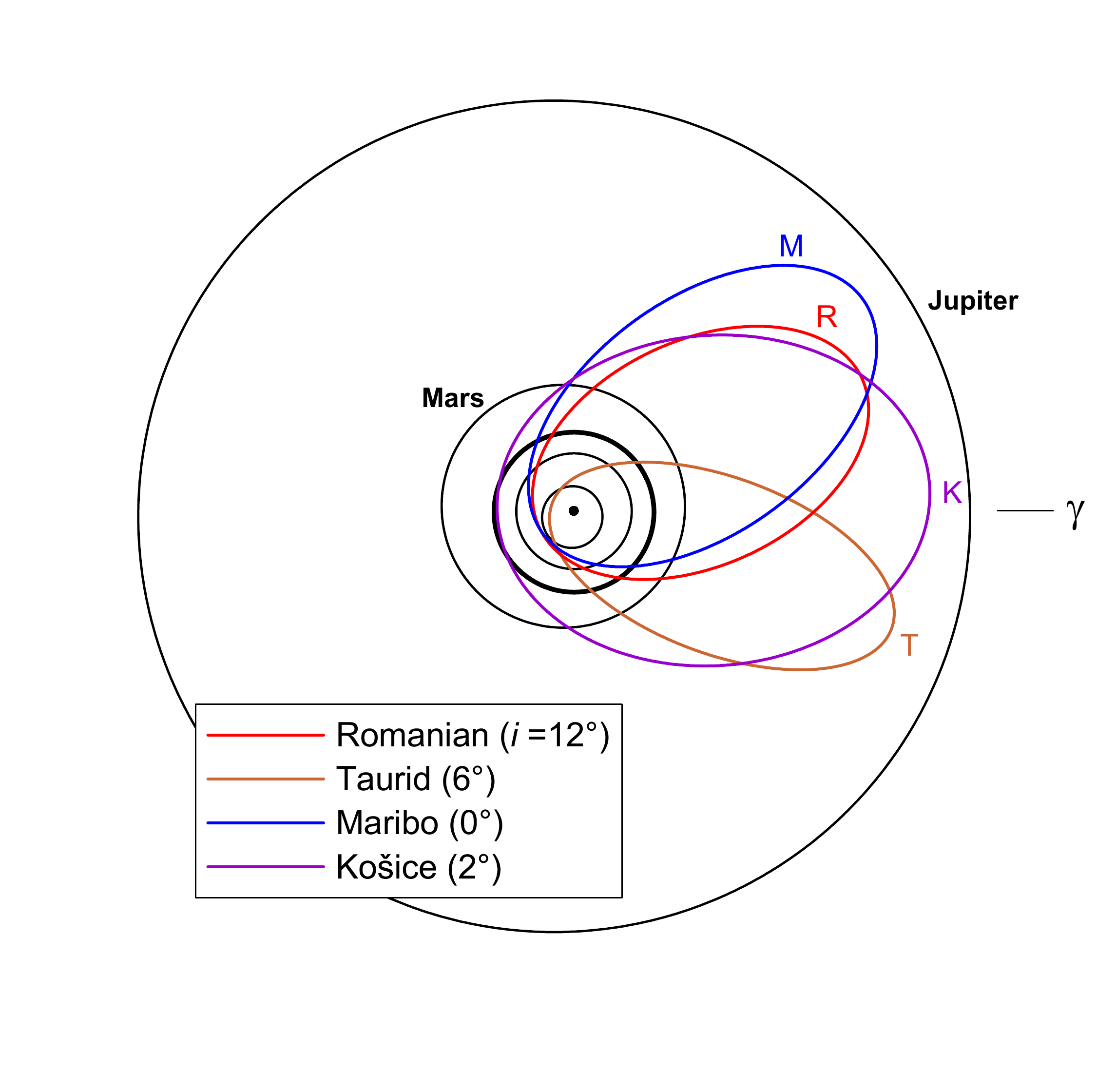} 
\caption{Heliocentric orbits (projected into the plane of ecliptic) of four superbolides. Orbit inclination is given
in the legend. The vernal equinox is to the right.}
\label{4orbits}
\end{figure}

In order to evaluate how typical or exceptional the behavior of the Romanian superbolide was, we will compare it
with three other superbolides with comparable brightness, for which we have radiometric light curves and 
fragmentation model could be constructed. 
The list of all compared bolides and their entry parameters are given in Table~\ref{4bolides}. 
Figure~\ref{4curves} shows calibrated radiometric light curves of all four bolides and Fig.~\ref{4orbits} 
shows their heliocentric orbits. We first describe the events individually and
then compare them.

\subsection{Ko\v{s}ice}

The Ko\v{s}ice meteorite fall occurred in Slovakia on February 28, 2010. The meteorites are ordinary 
chondrites of type H5, i.e.\ a very common type of stony meteorites. The bolide was studied in detail 
by \citet{Kosice}. The trajectory and orbit was determined on the basis of three casual video records. 
Radiometric curve from distant EN stations was also available and was used for modeling 
atmospheric fragmentation. We take here the results of \citet{Kosice}.
The maximum magnitude $-18$ was reached relatively high (at 36 km) but the
end height of only 17.4 km corresponds to strong meteoritic material.

\subsection{Maribo}

The Maribo meteorite fall occurred in Denmark on January 17, 2009. The meteorite is a carbonaceous 
chondrite of type CM, i.e.\ one of the softest known types of meteorites with density significantly lower 
than for ordinary chondrites \citep{Consolmagno}. The meteorite and the circumstances of its fall were 
described by \citet{MariboHaack}. The bolide trajectory, velocity, and orbit were estimated by several 
authors \citep[see][]{MariboSchult}. The results of an independent analysis of one casual video from Sweden, 
one distant all-sky image from the Netherlands, radar data from Germany and radiometric light curves from seven Czech stations
have been included in \citet{AIV}. 
All authors agree that the initial velocity was close to 28 km s$^{-1}$. It was surprising that 
meteorite of such a fragile type survived the atmospheric entry with such a high entry velocity. The maximum
magnitude $-19$ was reached at a height 37 km and the end height was 30.5 km.

The radiometric light curve and the fragmentation model were not yet published and we present them 
here for the first time. 

\subsection{Taurid bolide of October 31, 2015}

We selected for the comparison also a very bright Taurid bolide, which occurred over northern Poland 
on October, 31, 2015, at 18:05:20 UT. The bolide was studied by \citet{Olech} and independent 
analysis using EN data including radiometric curves was performed by Spurn\'{y} et al. (in preparation). 
We present here the light curve and fragmentation model. The initial mass was slightly lower than
 in other cases but still well above 1000 kg. The maximum magnitude was $-18.6$, so the bolide falls into
superbolide category.
The maximum height (81 km) and end height (58 km) are 
very large and the event is a good example of the entry of an extremely fragile cometary body, 
which was completely destroyed at high altitudes and no meteorite reached the ground. The bolide belonged to the
Taurid meteor shower. The parent body of this shower is comet 2P/Encke with the most probable
density of 800 kg m$^{-3}$ \citep{Encke}.

\begin{figure}
\includegraphics[width=\linewidth]{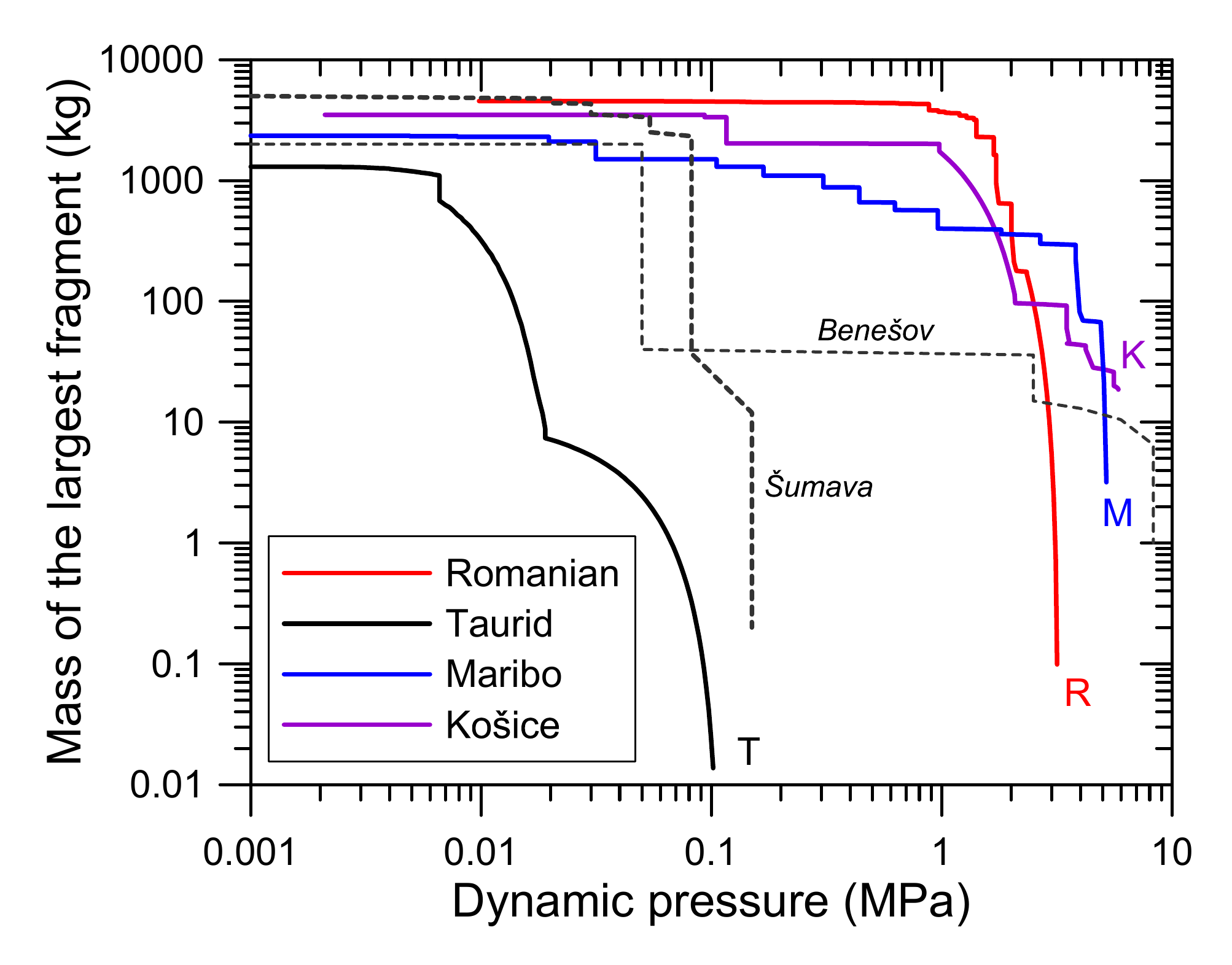} 
\caption{Modeled mass of the largest remaining fragment as a function of increasing dynamic pressure during the
atmospheric entry.}
\label{press-mass}
\end{figure}

\subsection{The comparison}

First we note that the orbits of all four bolides (Fig.~\ref{4orbits}) can be characterized by low inclinations
and aphelia between 4--4.5 AU.  Based on the orbits only, significant differences in material type
and, consequently, atmospheric behavior, could not be expected. Perhaps, the only indication of
different origin could be the somewhat lower eccentricity of the Ko\v{s}ice H5 chondrite.

The light curves in Fig.~\ref{4curves} have different character but light curves are influenced
by entry speeds and trajectory slopes, which varied from case to case (Table~\ref{4bolides}).
To reveal different material properties we have to compare fragmentation behavior in the atmosphere.
A good illustration of the fragmentation process is plotting the mass of the largest remaining fragment 
as a function of dynamic pressure acting on the body. Dynamic pressure, given by $p=\rho v^2$, where
$\rho$ is density of the atmosphere and $v$ is velocity, describes the actual mechanical action
of the atmosphere on the body. Using dynamic pressure as the independent variable enables us to
compare bolides with different slopes and speeds.
In fact, the actual dynamic pressure is $\Gamma \rho v^2$, where $\Gamma$ is the drag coefficient.
Since the value of $\Gamma$ is not known, but is probably similar in all cases (about 0.5), we use $\rho v^2$
for simplicity.

We can see in Fig.~\ref{press-mass} that the Taurid body started to loose the mass intensively under very
low dynamic pressures $<$0.01 MPa and was completely destroyed at 0.1 MPa. It was evidently
much more fragile than the other three bodies, though its strength was still much larger than 25 Pa expected for a rubble pile \citep{SS}.
The Romanian body lost
only negligible amount of mass before the dynamic pressure reached 0.9 MPa. In that respect
it was even more resistant than Ko\v{s}ice, which lost almost half of mass at 0.1 MPa. 
Maribo was loosing mass in repeating fragmentation events. At the time when dynamic pressure
reached 0.9 MPa, about 75\% of Maribo mass was lost.

After the dynamic pressure exceeded 1 MPa, the Romanian body started to disrupt severely.
 No fragment larger than 10 grams probably survived pressures above 3 MPa. Although Ko\v{s}ice also 
disrupted at 1 MPa, the difference is, that many macroscopic fragments survived the disruption
and some large fragments ($\geq 10$ kg) survived the maximal pressure of almost 6 MPa. 
They were then decelerated and dynamic pressure therefore started to decrease, while ablation, 
and possibly also some fragmentation, continued. 
The largest recovered meteorite had a mass of 2.37 kg \citep{Kosice, KosiceToth}.
Maribo fragmented most severely at 3--4 MPa (the exact value is uncertain because of lack of
detailed deceleration data). The mass of the only recovered meteorite was 26 g and we do not
think that any meteorites of significantly larger mass landed.

We also included in Fig.~\ref{press-mass} two older superbolides observed by EN cameras, namely the
$-21.5$ magnitude \v{Sumava}  (December 4, 1974)  and the $-19.5$ magnitude Bene\v{s}ov (May 7, 1991) \citep{icarus96}. 
Since radiometric light curves are not available in these cases, detailed fragmentation models could not be constructed.
For \v{Sumava} we plotted the original analysis of \citet{icarus96}. For Bene\v{s}ov, the recent re-evaluation of the
early-stage fragmentation \citep{IAUS318} was used. 

\v{Sumava}, also a member of the Taurid complex,
was a fragile body. As it can be seen in Fig.~\ref{press-mass}, it resisted fragmentation little bit more than the
EN\,311015 Taurid, nevertheless,  was completely destroyed at dynamic pressure of the order of 0.1 MPa.
Bene\v{s}ov was a deeply penetrating bolide and dropped stony meteorites, which, surprisingly, were of different
mineralogical types \citep{SpurnyBenesov}. The body disrupted into small pieces ($\sim$ 40 kg) at even lower pressure than \v{S}umava
but the small pieces survived until pressures 2--9 MPa. Both meteorite recoveries and atmospheric behavior
suggest that  Bene\v{s}ov was a weakly bound conglomerate of intrinsically relatively strong components.
It was probably a product of a collision of asteroids of different compositions.

Ko\v{s}ice was an example of a cracked body made from an intrinsically strong material.
The Romanian body behaved differently. There were apparently no large scale failures (cracks).
Most of the body remained intact until 1 MPa. After that, however, the body disrupted into
quite small fragments or dust. It seems that the intrinsic material strength was of the
order of 1 MPa. This is almost two orders of magnitude lower strength than for pristine
(uncracked) ordinary chondrites \citep{Popova} and almost two orders of magnitude  higher strength than 
for material from comet Encke. One may speculate that the Romanian body was
an carbonaceous asteroid. Maribo, however, showed different behavior with many separate
fragmentation events over a wide range of dynamic pressures from $\sim 5$ kPa 
(where only 1\% of mass was lost; more severe fragmentations started at 20--30 kPa)
to a few MPa.
That behavior suggests a hierarchical structure. It seems more likely that the Romanian bolide represents a different type of
asteroidal material, which is not represented in meteorite collections. It is weak but structurally homogeneous.

\section{Discussion}

\subsection{US Government sensor data}

We were able to provide reliable trajectory and velocity data for the January 7, 2015, superbolide observed over Romania.
This superbolide was detected also by the US Government sensors (USGovS). The data were used by
\citet{Brown} in their study of meter-scale impactors. Unfortunately, the speed reported by the USGovS (35.7
km s$^{-1}$) was by 8 km s$^{-1}$ (almost 30\%) larger than the real speed. The derived
orbit was distinctly cometary with Tisserand parameter $T_J=1.75$ and aphelion at 9.2 AU. In reality, the
aphelion was just above 4 AU and the orbit can be classified as asteroidal. This example shows that big care must be taken
when using the USGovS data. 

We can compare also other quantities reported for this event at the USGovS bolide 
webpage\footnote{http://neo.jpl.nasa.gov/fireballs/} and in \citet{Brown}. The radiant position (R.A. = 119.8$^\circ$, Decl. = +7.0$^\circ$)
is off by seven degrees, which is also significant for orbit computation. Bolide location is given only
to one decimal point (45.7N, 26.9E) and is correct within that precision. The reported height of maximum (45.5 km)
is too high by 2.7 km. It lies, nevertheless, within the bright phase of the bolide (see Fig.~\ref{lc-model}). 
The total impact energy was reported to be 0.4 kt TNT. Our mass (4500 kg) and speed (27.76 km s$^{-1}$)
give $1.7\times10^{12}$ J = 0.41 kt TNT. Since our mass estimate is rather uncertain, this is a surprisingly  good agreement.

The bolide position and energy reported by the USGovS seem therefore to be more reliable than the velocity vector.
Note that Ko\v{s}ice bolide was observed by the USGovS as well. The speed agrees within 0.1
km s$^{-1}$ in this case and 
the radiant difference is only 4 degrees.

\subsection{Derivation of asteroid structure from superbolide data}

The main topic of this paper was the study of the structure and strength of small asteroids using superbolide data. 
We modeled high resolution radiometric light curves with the knowledge of bolide trajectories and speeds.
The aim was to reveal atmospheric fragmentation history. 
We have to note that the models are not unambiguous and the light curves could be 
probably reproduced with different sets of parameters. Nevertheless, the main characteristics, especially the
heights of mass loss events and the amount of mass lost in form of small quickly evaporating fragments, can be revealed well. 
These characteristics are sufficient to study structural differences among impacting asteroids. We cannot be sure about 
the number and masses of macroscopic fragments. For that purpose detailed video records showing the motion of
individual fragments would be desirable. Such data are very rare. One example was the Mor\'{a}vka meteorite fall 
\citep{Moravka}, where, however, sufficiently good light curve was missing. It could be partly supplemented by good seismic
data, which also contain information about fragmentation events. 

Unfortunately, data available for most superbolides are very limited.  \citet{Brown} used 
the height of maximum as a proxy of asteroid strength. Our small sample shows that bodies with quite different material properties
can have similar peak height. Peak height is therefore
poor discriminator of material strength. Extremely fragile (cometary) or very strong bodies can be distinguished but nothing can
be said about the majority of events with intermediate peak heights.

The end height is a better discriminator (Ko\v{s}ice 17 km, Maribo 31 km, Romanian event 39 km if neglecting the most detailed video,
Taurid 58 km). However, end height is more dependent on observational
techniques and circumstances and is not available at all in the USGovS data.
The classical PE criterion \citep{PE} is based on end height, although the entry mass, speed, and angle must be also taken into account.
This criterion was, nevertheless, developed for ordinary bolides and may lead to misleading results when applied to superbolides.
The Romanian event would be classified as type IIIB, i.e.\ soft cometary material, using the PE criterion, even if the end height of 36 km
from the most detailed video were used. 
The Taurid would be also IIIB, while Maribo and Ko\v{s}ice fall into type II.
Note that the original luminous efficiency of \citet{PE} must be used when computing the
entry mass for the PE formula \citep[it can be found also in][]{Ceplecha88}. The modern luminous efficiency for bolides is higher 
\citep{ReVelle2001}.
\citet{Ceplecha94} used the old luminous efficiency and overestimated the masses and sizes of the studied meteoroids.

Bolide end heights were recently discussed also by \citet{Moreno}. 
Their methodology, however, relies on observed deceleration along the trajectory
and does not take into account bolide luminosity, and is therefore not suitable in the present cases.

\section{Conclusion}

The main conclusion of this paper is the large structural diversity of meter-sized asteroids. Although many types of 
meteorites are known,  from the point of view of structure and density, there are just three main types: metals, silicate-rich 
stones, and carbonaceous stones. Of course, asteroids of cometary origin with low density and strength are also
expected to exist. The Romanian event, however,  probably did not belong to any of these types. The bulk strength
and the material strength were both of the order of 1 MPa. We expect the material to have high microporosity
but not so high as cometary materials. Judging from the orbit, the Romanian asteroid probably
originated in the middle or outer asteroid belt.

The implication of large structural diversity of small asteroids is difficulty of predicting impact consequences
for an asteroid of known size but unknown material properties.

\begin{table*}
\caption{List of supplementary video files}
\label{videofiles}
\begin{tabular}{llll}
\hline
File & Video site & Credit & Note\\ \hline
Cluj-Napoca.mp4 & Cluj-Napoca, C\^ampul P\^ainii str. &  SECPRAL COM \\
EforieSud-small.avi &Eforie Sud & High School ``Carmen Sylva'' & low resolution \\
EforieSud-3fps.avi  & " & " & low frame rate \\
Fierbinti.mp4 & Fierbinti, Nordului str. & Adrian Pascale & color\\
Gornovita.avi & Gornovita observatory & Marian Lucian Achim \\
Sibiu.avi & Sibiu Municipal Stadium & Sibiu City Hall \\
Voslobeni.avi & Vo\c{s}lobeni, Heveder str. & Balint Ede \\
\hline
\end{tabular}
\end{table*}

\section*{Acknowledgements}

We are grateful to the following people, who helped us to calibrate the videos by taking on site calibration images:  
Lucian Hudin, Attila Munzlinger, Radu Cornea, Florin Serbu, Marian Lucian Achim.
We thank Razvan Andrei, Radu Mihailescu, Costi Movileanu, Dan Zavoianu, Alexandru Badea, 
Adrian Sonka, Mihai Dascalu, and Elisabeta Petrescu for participating in meteorite searches.
This work was supported by the project No.\  16-00761S from GA\v{C}R,
Praemium Academiae of the Czech Academy of Sciences, and the Czech institutional project RVO:67985815.


\appendix

\section{Supplementary files}
\label{suppl}

The video records of the January 7 bolide used in our analysis are provided as supplementary files to the online version of this article . 
The list of the files is given in Table~\ref{videofiles}. Note that the files are meant mostly for illustrative
purposes. Although care was taken to preserve video quality, slight deterioration or deformation may have occurred
when converting between different formats and codecs. In case of Eforie Sud we were able to provide for technical reasons the video
in two limited versions only; one with lower resolution and one with limited number of frames.

In addition, we provide the derived coordinates (azimuths, zenith distances) of the bolide as seen from the individual
sites, and the radiometric light curve. Both are given as self-explanatory text files, \texttt{coordinates.txt} and \texttt{lightcurve.txt},
respectively.



\end{document}